%% file: main.tex
\documentclass[manuscript,authorversion,nonacm]{acmart}

\AtBeginDocument{%
  }

\setcopyright{acmlicensed}
\copyrightyear{2018}
\acmYear{2018}
\acmDOI{XXXXXXX.XXXXXXX}
\acmConference[Conference acronym 'XX]{Make sure to enter the correct
  conference title from your rights confirmation email}{June 03--05,
  2018}{Woodstock, NY}
\acmISBN{978-1-4503-XXXX-X/2018/06}

\usepackage[most]{tcolorbox}
\usepackage{subcaption}

\begin{document}

\title{The Work Behind Delegation: A Framework for Supervising AI Coding Agents}

\author{Yeon Su Park}
\orcid{0009-0004-3071-6664}
\affiliation{
  \institution{School of Computing, KAIST}
  \city{Daejeon}
  \country{Republic of Korea}
}
\email{yeonsupark@kaist.ac.kr}

\author{Nadia Arvi}
\orcid{0009-0001-9527-464X}
\affiliation{
  \institution{School of Computing, KAIST}
  \city{Daejeon}
  \country{Republic of Korea}
}
\email{nadia.arvi@kaist.ac.kr}

\author{Hae Ri Lee}
\orcid{0009-0000-8003-7265}
\affiliation{
  \institution{School of Computing, KAIST}
  \city{Daejeon}
  \country{Republic of Korea}
}
\email{harriet@kaist.ac.kr}

\author{Sehoon Lim}
\orcid{0009-0008-6179-2036}
\affiliation{
  \institution{School of Computing, KAIST}
  \city{Daejeon}
  \country{Republic of Korea}
}
\email{sehoon1106@kaist.ac.kr}

\author{Qianou Ma}
\orcid{0009-0002-8634-130X}
\affiliation{
  \institution{Human-Computer-Interaction Institute, Carnegie Mellon University}
  \city{Pittsburgh, Pennsylvania}
  \country{USA}
}
\email{qianoum@cs.cmu.edu}

\author{Juho Kim}
\orcid{0000-0001-6348-4127}
\affiliation{
  \institution{School of Computing, KAIST}
  \city{Daejeon}
  \country{Republic of Korea}
}
\affiliation{
  \institution{SkillBench}
  \city{Santa Barbara, CA}
  \country{USA}
}
\email{juhokim@kaist.ac.kr}

\renewcommand{\shortauthors}{Yeon Su Park et al.}

\begin{abstract}
    \input{sections/0_Abstract}
\end{abstract}



\begin{teaserfigure}
    \includegraphics[width=\textwidth]{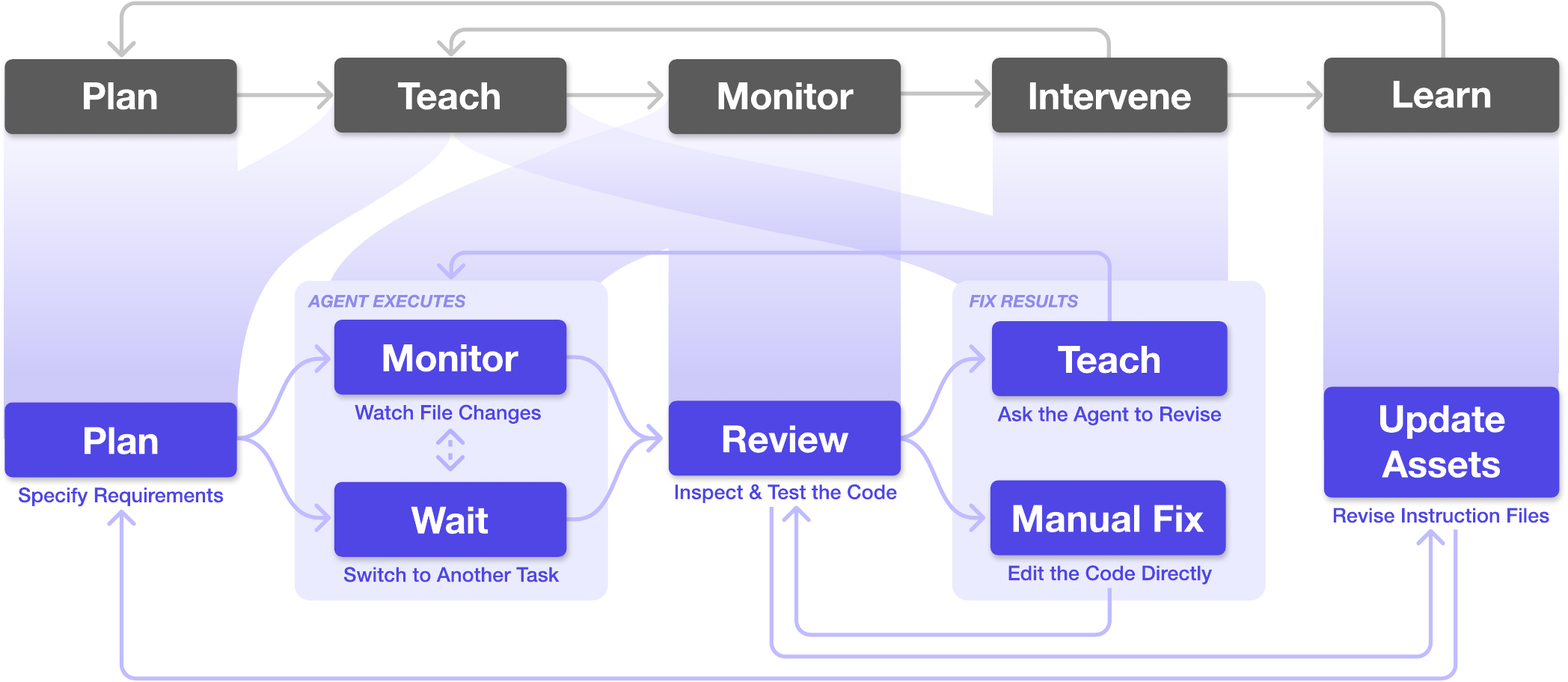}
    \caption{We introduce a new framework for supervising AI coding agents (bottom) by reconfiguring the five stages of Sheridan's classical human supervisory control (top). Sheridan's five stages—\textsc{Plan}, \textsc{Teach}, \textsc{Monitor}, \textsc{Intervene}, and \textsc{Learn}—map onto our seven stages—\textsc{Plan}, \textsc{Monitor}, \textsc{Wait}, \textsc{Review}, \textsc{Teach}, \textsc{Manual Fix}, and \textsc{Update Assets}—as shown by the connecting bands.}
    \Description{Two rows of boxes joined by pale vertical bands showing how the top row maps onto the bottom. Top row: five boxes—Plan, Teach, Monitor, Intervene, Learn—connected left to right by arrows in sequence, with two additional feedback arrows arcing back from Intervene to Teach and from Learn to Plan. Bottom row: seven boxes—Plan, Monitor, Wait, Review, Teach, Manual Fix, Update Assets—where Monitor/Wait and Teach/Manual Fix are each grouped as a pair. The main path runs Plan to Monitor/Wait to Review, then to either Teach or Manual Fix, looping back to Monitor/Wait or Review respectively; an arrow also connects Review to Update Assets, and another arrow loops back from Update Assets to Plan. The connecting bands: top Plan maps to bottom Plan; top Teach maps to bottom Plan and to Teach/Manual Fix; top Monitor maps to Monitor/Wait and to Review; top Intervene maps to Teach/Manual Fix; top Learn maps to Update Assets.}
    \label{fig:teaser}
\end{teaserfigure}

\received{20 February 2007}
\received[revised]{12 March 2009}
\received[accepted]{5 June 2009}

\maketitle

\input{sections/1_Introduction}
\input{sections/2_Related_Work}
\input{sections/3_Background}
\input{sections/4_Method}
\input{sections/5_Framework}
\input{sections/6_Application}
\input{sections/7_Findings}

\input{sections/8_Discussion}
\input{sections/9_Limitation}
\input{sections/10_Conclusion}


\bibliographystyle{ACM-Reference-Format}
\bibliography{references}

\appendix
\input{sections/99_Appendix}

\end{document}

%% file: sections/0_Abstract.tex
As AI coding agents carry out development tasks with greater autonomy, developers are shifting from direct implementation toward supervising delegated work. Yet existing research offers limited understanding of how developers organize supervisory activities into connected workflows. Drawing on observations and workflow diagrams from 19 experienced developers, we reconfigure Sheridan’s framework of human supervisory control into seven stages and the connecting loops for supervising AI coding agents. We applied the framework to public developer discussions on Reddit and found that supervisory demands extend across stages and that developers manage them by concentrating effort in planning, delegating supervisory work to other agents, and turning recurring guidance into reusable assets. Our framework provides a useful analytical lens for understanding how developers supervise AI coding agents by capturing how supervision is structured in agentic software development. 

%% file: sections/1_Introduction.tex
\section{Introduction}

AI coding tools have rapidly expanded their capabilities, moving from assisting developers \cite{santos2025llmcoding} to carrying out software development tasks with greater autonomy~\cite{mozannar2024cups, robbes2026agentic}. With the emergence of AI coding agents such as Claude Code~\cite{claudecode2025} and Codex~\cite{openaicodex2025}, developers can now provide high-level direction and leave the execution to the agent. Given a task description, agents can autonomously plan their own approach, explore the codebase, run commands, and make changes across multiple files~\cite{dong2025surveycodegen, guo2025surveybenchmarkssolutions}. As developers delegate more of this execution to agents, their role shifts from direct implementation toward coordination and oversight~\cite{alenezi2026rethinkingagentic}. Therefore, developers increasingly act as \textit{supervisors} of AI coding agents, setting the direction of delegated work, tracking progress, and evaluating its output~\cite{sapkota2025vibevsagentic}.

As developers take on this supervisory role, understanding what effective supervision of AI coding agents requires becomes important. Prior work has examined what developers do at particular points in the process, such as providing clear context and explicit instructions~\cite{huang2025professional}, asking agents to explain their changes~\cite{baumann2026swechat}, and providing corrective feedback when execution goes wrong~\cite{chen2026code}. However, supervision extends across a development task, and effort invested at one point can affect what needs to be monitored, corrected, or reviewed at subsequent points in the task~\cite{dhanorkar2026human, zhou2026should, chen2026comparing}. Given limited human attention and effort, these dependencies highlight the need to consider supervision as a connected process rather than as a set of separate activities. Understanding this process requires an analytical approach that captures how supervisory activities are organized and connected across a task.

Earlier research on automation provides a useful foundation for understanding how human roles change as task execution is increasingly automated by systems such as aircraft autopilots and industrial robots~\cite{parasuraman2000model, sheridan2012hsc}. Among this body of work, we build on Sheridan's framework of \textit{human supervisory control}, which conceptualizes supervision as a process organized around five stages: \textsc{Plan}, \textsc{Teach}, \textsc{Monitor}, \textsc{Intervene}, and \textsc{Learn}~\cite{sheridan1976general, sheridan1992telerobotics}. These stages are linked through recurrent loops, highlighting how one stage shapes subsequent supervisory demands. While Sheridan's framework was grounded largely in automated systems that executed human-specified goals or instructions, AI coding agents have greater autonomy in determining how to accomplish delegated tasks. Hence, it remains unclear whether Sheridan's supervisory framework can fully capture supervision under this more agentic form of delegation.

This motivates a central question for our work: \textit{How do developers organize supervision over the course of a task delegated to AI coding agents?} To answer this, we developed a framework for supervising AI coding agents by reconfiguring Sheridan's framework (Figure~\ref{fig:teaser}). The framework was based on a qualitative study with 19 experienced developers, where we observed participants working with agents and asked them to reconstruct their supervision workflows as diagrams. Our analysis identified seven supervisory stages---\textsc{Plan}, \textsc{Monitor}, \textsc{Wait}, \textsc{Review}, \textsc{Teach}, \textsc{Manual Fix}, and \textsc{Update Assets}---and loops connecting them. Together, these stages and loops capture how supervision is organized in agentic software development, providing an analytic lens for studying developer supervision of AI coding agents.

We then applied the framework to developer discussions on Reddit to examine supervisory patterns across the development process. We identified three demands that span multiple stages: maintaining alignment between agent behavior and developer intent, enforcing quality beyond task completion, and maintaining understanding of agents' work. We also found three strategies for managing these demands: concentrating effort in planning, distributing supervisory work to other agents, and externalizing repeated supervision into reusable assets. Through the framework, we show how supervisory demands extend across stages and how developers redistribute their effort in response.

The contributions of this paper are as follows:
\begin{itemize}
    \item A framework for supervising AI coding agents that consists of seven stages and the loops connecting them.
    \item Findings from applying the framework to examine supervisory demands and the strategies developers use to manage them across the development process.
\end{itemize}

%% file: sections/2_Related_Work.tex
\section{Related Work}

We review prior work to contextualize our framework for developer supervision of AI coding agents: (1) AI-assisted software development, (2) frameworks for human--agent collaboration, and (3) human roles in the agentic era.

\subsection{AI-Assisted Software Development}
AI assistance in software development has shifted from providing localized code suggestions~\cite{sarkar2022programwithai, prather2023s, fan2026when, ma2023pair} toward increasingly autonomous AI agents that can carry out development tasks with limited human intervention~\cite{liu2026agentssurvey, fawzy2025vibecodinglitreview, ray2025reviewvibecoding}. As AI systems have become more capable~\cite{dong2025surveycodegen, feng2025levels}, how developers interact with them has also changed.

Prior work has examined AI assistance in software development and its integration into developer workflows. Early studies focused on developers' interactions with individual AI-generated suggestions, examining how they evaluated, accepted, and modified suggestions during programming~\cite{barke2023groundedcopilot, mozannar2024cups}. More recent work has expanded attention to AI use across programming sessions and broader development workflows, drawing on interviews with practitioners~\cite{ullrich2025requirements, dhanorkar2026human}, interaction logs from real-world use~\cite{wu2026how, tang2026programming}, and longitudinal telemetry~\cite{sergeyuk2026evolving}. Together, these studies highlight the importance of understanding developer--AI interaction as AI becomes deeply integrated into software development.

Agentic AI introduces new dynamics of developer--AI interaction. AI agents can proactively plan and execute multi-step actions~\cite{chen2026code}, incorporate broader contextual information~\cite{guo2025surveybenchmarkssolutions}, and produce artifacts across development workflows~\cite{zamfirescu-pereira2025beyondcode}. As developers delegate more execution to agents, supervision becomes increasingly important, yet its organization across a development task remains underexplored. Therefore, in this work, we examine the supervisory activities developers perform and how these activities connect throughout the process.

\subsection{Frameworks for Human--Agent Collaboration}
A growing body of empirical work has examined human--agent collaboration through interviews~\cite{dhanorkar2026human}, observational studies~\cite{huang2025professional, shome2026johnny}, and real-world interaction logs~\cite{tang2026programming, baumann2026swechat}. Building on these findings, researchers have developed different conceptualizations of human--agent interaction, capturing recurring modes, activities, and dimensions of collaboration. For example, \citet{barke2023groundedcopilot} distinguish between exploration and acceleration modes of interaction with programming assistants, while \citet{dhanorkar2026human} identify four forms of oversight: a priori control, co-planning, real-time monitoring, and post-hoc review. Other studies describe human--agent interaction in terms of the degree of human involvement~\cite{chen2026code}, time allocation across activity states~\cite{mozannar2024cups}, and the distribution of initiative between human and system~\cite{maier2026partnering}.

Beyond these study-specific conceptualizations, recent work has proposed more explicit frameworks for human--agent collaboration. \citet{feng2025levels} structure collaboration around levels of agent autonomy and corresponding user roles, while \citet{wang2025interaction} propose a process-oriented framework that connects interaction, process, and infrastructure. These frameworks provide systematic ways to structure human--agent collaboration, but offer limited guidance to collaboration centered on supervising delegated agent work. In this work, we develop a framework that captures the supervisory activities involved and how they connect across a development task.

\subsection{Human Roles in the Agentic Era}
The emergence of AI as a collaborator is not unique to software development. Similar shifts toward human--AI collaboration have been observed across domains such as design~\cite{xu2025productive, cavallin2026designers}, writing~\cite{lehmann2026collaborative}, and knowledge work~\cite{yun2025generative}, where AI systems have become integrated into everyday workflows. As these systems become more agentic and take on greater responsibility for execution, human roles are being reconceptualized. Rather than acting as primary executors while AI provides passive assistance, humans are increasingly shifting toward supervisory roles over delegated work by defining goals~\cite{feng2026cocoa, ullrich2025requirements, ma2025prompt}, evaluating outcomes~\cite{feng2026juniortosenior}, and intervening when necessary~\cite{zhou2026should}.

These shifts also align with longstanding perspectives in automation research, which describe how responsibilities shift between humans and systems as system autonomy increases~\cite{parasuraman2000model, sheridan2012hsc}. Taken together, these perspectives point toward a shift in human contribution from direct execution toward higher-level activities such as providing context, exercising judgment, and guiding AI-generated work~\cite{feng2026juniortosenior, Cheruvu2025}. Building on these observations, we examine how this shift manifests in software development as developers increasingly supervise work delegated to AI agents.

%% file: sections/3_Background.tex
\section{Background} \label{sec:background}

\begin{figure*}[hbt!]
    \centering
    \includegraphics[width=0.85\linewidth]{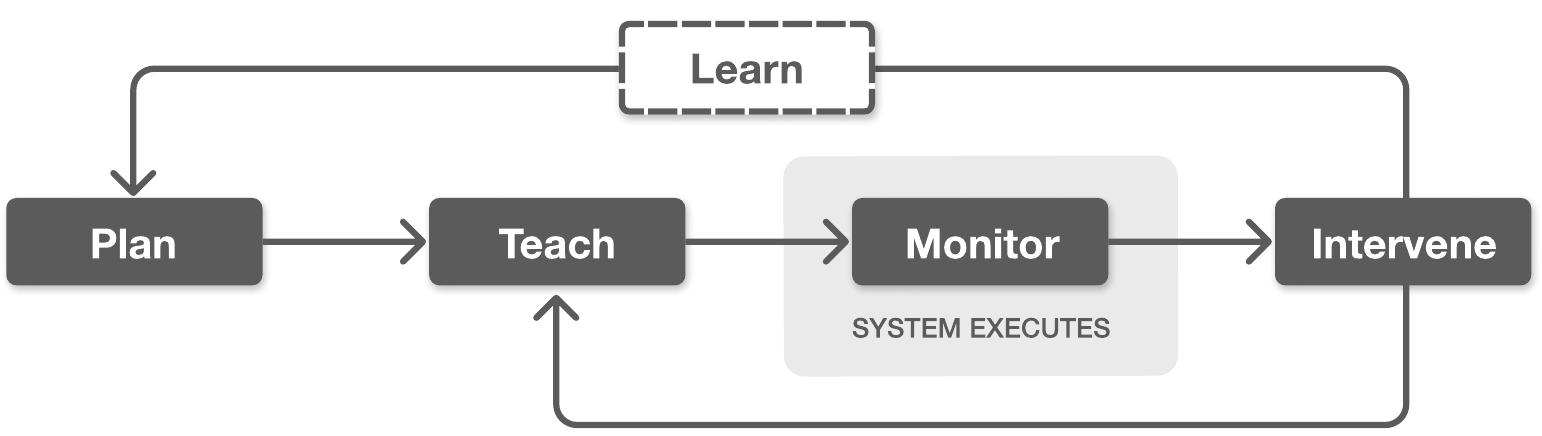}
    \caption{Sheridan's framework of human supervisory control, consisting of five stages---\textsc{Plan}, \textsc{Teach}, \textsc{Monitor}, \textsc{Intervene}, and \textsc{Learn}---connected in sequence, with an inner loop from \textsc{Intervene} to \textsc{Teach} and an outer loop from \textsc{Intervene} through \textsc{Learn} to \textsc{Plan}.}
    \label{fig:sheridan}
    \Description{A flow diagram with a row of boxes---Plan, Teach, Monitor, Intervene---connected left to right by arrows in sequence, where Monitor sits inside a shaded region labeled ``system executes''. An arrow loops from Intervene back down to Teach. Above the row, a dashed box labeled Learn is connected by two arrows: one arrives from Intervene, and one leaves toward Plan.}
\end{figure*}

In this section, we provide background on Sheridan's \textit{human supervisory control} framework, which serves as the theoretical starting point for our work examining how developers supervise AI coding agents. The \textit{human supervisory control} framework describes how a human supervisor directs and oversees automated systems such as aircraft autopilots and industrial robots~\cite{sheridan1992telerobotics, sheridan1976general, sheridan2012hsc}. The framework consists of five supervisory stages: \textsc{Plan}, \textsc{Teach}, \textsc{Monitor}, \textsc{Intervene}, and \textsc{Learn}, connected by recurring loops as illustrated in Figure~\ref{fig:sheridan}. We briefly explain each stage below.

The supervisor begins with \textsc{Plan}, where they develop an understanding of the process and system capabilities, then determine the goals, constraints, and overall strategy. In \textsc{Teach}, these plans are translated into instructions for the automated system. During execution, \textsc{Monitor} involves observing the system and assessing whether it is progressing as intended, while \textsc{Intervene} occurs when the supervisor needs to revise the instructions or take more direct control. Revising instructions returns the process to \textsc{Teach}, whereas satisfactory completion leads to \textsc{Learn}, where the supervisor reflects on key events and prior decisions to inform future \textsc{Plan} activities. Together, these stages form recurring control loops in which supervision can return to earlier stages as execution progresses (Figure~\ref{fig:sheridan}).

%% file: sections/4_Method.tex
\section{Method}

We conducted a qualitative study with experienced software developers to develop a framework for supervising AI coding agents. We used Sheridan's framework of human supervisory control~\cite{sheridan1976general, sheridan1992telerobotics} (\S\ref{sec:background}) as a theoretical starting point to examine how its stages and loops translate to the supervision of AI coding agents as developers delegate more implementation to them. We observed participants working on their own coding tasks with their usual agent setups and asked them to reconstruct their supervision workflows as diagrams. We analyzed these workflow diagrams together with task observations and think-aloud data to develop our framework, which we introduce in \S\ref{sec:framework}. 

In this section, we describe how we designed and conducted the study in detail. The study was reviewed and approved by the Institutional Review Board (IRB) of our institution prior to conducting the study.

\subsection{Participants}
We distributed a screening survey through university forums and developer communities in South Korea, as well as LinkedIn groups and networks for international recruitment. The survey collected information about respondents' software development background, AI coding agent use (i.e., which agents they used, how frequently they used them, and the types of tasks they delegated), and the coding task they could bring to the study session.

Based on the survey responses, we applied three eligibility criteria: (1) at least three years of professional software development or graduate-level research experience in a software-intensive field (e.g., computer science, electrical engineering), (2) regular use of AI coding agents for substantive development tasks in professional, research, or other real-world projects, and (3) the ability to bring a shareable coding task with a clearly defined goal for a 40-minute session. These criteria ensured that participants had sufficient development experience to judge agent output critically and sufficient familiarity with agents to have established supervisory practices of their own.

We recruited 20 participants who met these criteria. One participant was unable to use their regular agent setup during the session and was excluded from the analysis, resulting in a final sample of 19 participants. The full list of participants and their information is shown in Table~\ref{tab:participants}. We deliberately recruited participants across diverse roles and development backgrounds to capture variation in how developers supervise AI coding agents.

Sessions were conducted in Korean or English, and Korean transcripts were translated into English for analysis. All eligible participants received either a 120 USD Amazon gift card or 150,000 KRW ($\approx$ 98 USD) for a 90-minute study.

\renewcommand{\arraystretch}{1.25}
\begin{table*}[t!]
\small
\caption{Overview of the 19 participants and their study sessions. In the Agent Setup column, ``+'' denotes an integrated setup, ``\&'' denotes two agents used independently, and ``in'' denotes an agent running within an orchestration environment.}
\label{tab:participants}
\begin{tabular}{c|c|c|l|l}
\toprule
\textbf{ID} & \textbf{Current Role}                                                               & \textbf{\begin{tabular}[c]{@{}c@{}}Experience\\ (Years)\end{tabular}} & \multicolumn{1}{c|}{\textbf{Agent Setup}} & \multicolumn{1}{c}{\textbf{Task}}                  \\ \hline
P1          & Frontend Developer                                                                  & 26                                                                    & GitHub Copilot                            & Adding recommendation comparison pages             \\ \hline
P2          & \begin{tabular}[c]{@{}c@{}}Graduate Student\end{tabular} & 4                                                                     & Claude Code                               & Generating publication-ready research figures      \\ \hline
P3          & Backend Developer                                                                   & 5                                                                     & Codex + Lazy Codex                        & Building a web game prototype                      \\ \hline
P4          & Product Manager                                                                     & 6                                                                     & Claude Code + Superpowers                 & Pivoting a scheduler into a contact manager \\ \hline
P5          & \begin{tabular}[c]{@{}c@{}}Graduate Student\end{tabular}       & 3                                                                     & Codex + OhMyCodex                         & Refining a 3D simulation app UI and docs           \\ \hline
P6          & Backend Developer                                                                   & 6                                                                     & Codex                                     & Fixing trading backend order bugs                  \\ \hline
P7          & Full Stack Developer                                                                & 5                                                                     & Claude Code in Cmux                       & Adding tests to a web service                      \\ \hline
P8          & \begin{tabular}[c]{@{}c@{}}Graduate Student\end{tabular}       & 4                                                                     & Codex                                     & Adding NPC movement and dialogue                   \\ \hline
P9          & Full Stack Developer                                                                & 3                                                                     & OpenCode + OhMyOpenCode                   & Adding diagram support to a math generator         \\ \hline
P10         & Indie Game Developer                                                                & 5                                                                     & Codex + Unity MCP                         & Adding currency UI and sound effects               \\ \hline
P11         & Backend Developer                                                                   & 7                                                                     & Codex                                     & Building a chess mistake-analysis website          \\ \hline
P12         & \begin{tabular}[c]{@{}c@{}}ML Engineer\end{tabular}                 & 7                                                                     & Claude Code in Conductor                  & Generating labels for video augmentation pairs     \\ \hline
P13         & Data Analyst                                                                        & 5                                                                     & Codex                                     & Adding mood-based recommendations                  \\ \hline
P14         & Frontend Developer                                                                  & 5                                                                     & Codex + Claude Code                       & Adding an Android reflection widget                \\ \hline
P15         & \begin{tabular}[c]{@{}c@{}}Graduate Student\end{tabular}      & 5                                                                     & Hermes \& Claude Code                     & Optimizing load time and accessibility             \\ \hline
P16         & Full Stack Developer                                                                & 3                                                                     & Claude Code                               & Building a bid-document analysis website           \\ \hline
P17         & Full Stack Developer                                                                & 6                                                                     & Claude Code                               & Migrating a blog backend to PostgreSQL             \\ \hline
P18         & Full Stack Developer                                                                & 6                                                                     & Claude Code                               & Adding dashboard and delete features               \\ \hline
P19         & Founding Engineer                                                                   & 15                                                                    & Antigravity                               & Migrating a remote-control bridge and CLI         
\\ \bottomrule
\end{tabular}
\Description{A table of 19 participants with their ID, current role, years of experience, agent setup, and study task.}
\end{table*}

\subsection{Task}
Following \citet{huang2025professional}, participants were asked to bring their own task from their ongoing project rather than complete a given standardized task. This allowed us to observe realistic development workflows~\cite{becker2025measuring}, while introducing natural variation in tasks, codebases, and agent configurations across the study sessions~\cite{ko2015practical}.

Each task needed a clearly defined goal that could be pursued within approximately 40 minutes. This scope allowed us to observe supervision end-to-end within a single session. Tasks could come from ongoing work, open-source projects, side projects, or research code, as long as they could be shared with the research team and did not contain confidential or personal information. The \textit{Task} column in Table~\ref{tab:participants} summarizes the task brought by each participant.

\subsection{Procedure}
We conducted the study remotely over Zoom\footnote{\url{https://zoom.us/}}. Each session lasted approximately 90 minutes and consisted of three phases: (1) a pre-task interview, (2) a think-aloud coding task, and (3) a supervision workflow modeling activity. All sessions were recorded and transcribed for later analysis. We describe each phase below.

\subsubsection{Pre-Task Interview}
We began with a semi-structured interview that expanded on participants' initial survey responses about their development background and agent use. Participants then introduced their prepared task and the concrete goal they intended to pursue during the session. We provided brief instructions on the think-aloud protocol as needed. This phase lasted approximately 10 minutes. The full interview questions are provided in Appendix~\ref{app:interview_questions}.

\subsubsection{Coding Session}
In this phase, participants worked on their prepared task in their own environment, sharing their screen(s) throughout. They were asked to think aloud as they worked, and we prompted them when they remained silent for an extended period. Sessions ended when participants felt they had reached their goal, or after approximately 40 minutes had passed. In the latter case, we briefly asked how much work remained.

\subsubsection{Workflow Design Session}
Participants then drew a diagram of their supervision workflow on a shared online whiteboard. To examine how human supervisory control translates to AI coding agents, we first introduced Sheridan's framework to participants. We emphasized that participants did not need to follow Sheridan's stages and asked them to construct their diagrams based on their own supervision practices. The activity proceeded in three steps.

First, participants reflected on the task they had just completed and divided their supervision into stages they considered meaningful and distinct. If a stage was difficult to name concisely, they could write a short description instead. Second, they arranged the stages in order and marked recurring transitions as loops. Finally, they considered how the diagram might extend to their other tasks and to other developers, revising it accordingly. As in the coding session, participants were asked to think aloud during the activity, explaining the meaning of each stage and loop and the reasoning behind it. This phase lasted approximately 30 minutes. Example diagrams are provided in Appendix~\ref{app:workflow_diagrams}.

\subsubsection{Inductive Content Analysis}

We analyzed participants' workflow diagrams, task observations, and think-aloud data following the content analysis approach of ~\citet{elo2008qualitative}. We inductively derived stages and loops in participants' supervision processes and used the findings to reconfigure Sheridan's framework for AI coding agents.

After the first eight sessions, the three authors agreed that common patterns were emerging across participants. Each author independently reviewed all eight participants' data to familiarize themselves with the data. We then collaboratively coded each participant's supervision process across the three sources. The diagrams provided participants' own representations of their processes, while task observations and think-aloud data clarified ambiguities and captured omitted steps. Through iterative comparison, we grouped codes representing similar supervisory functions, aligned their terminology and granularity, and organized them into an initial framework of stages and loops.

We first checked the framework against the first eight cases. After each of the remaining 11 sessions, we compared the participant's workflow with the framework and examined whether stages, transitions, or loops required revision. When a mismatch arose, the three authors revisited the data, discussed the case, and revised the framework as needed. This process led to the removal of one stage and the addition of one loop. As no new stages or loops emerged in later sessions, we stopped data collection after 19 participants. The final framework consists of seven stages---\textsc{Plan}, \textsc{Monitor}, \textsc{Wait}, \textsc{Review}, \textsc{Teach}, \textsc{Manual Fix}, and \textsc{Update Assets}---and the loops connecting them. We present it in detail in \S\ref{sec:framework}.

%% file: sections/5_Framework.tex
\section{A Framework for Supervising AI Coding Agents} \label{sec:framework}

\begin{figure*}[hbt!]
    \centering
    \includegraphics[width=0.85\linewidth]{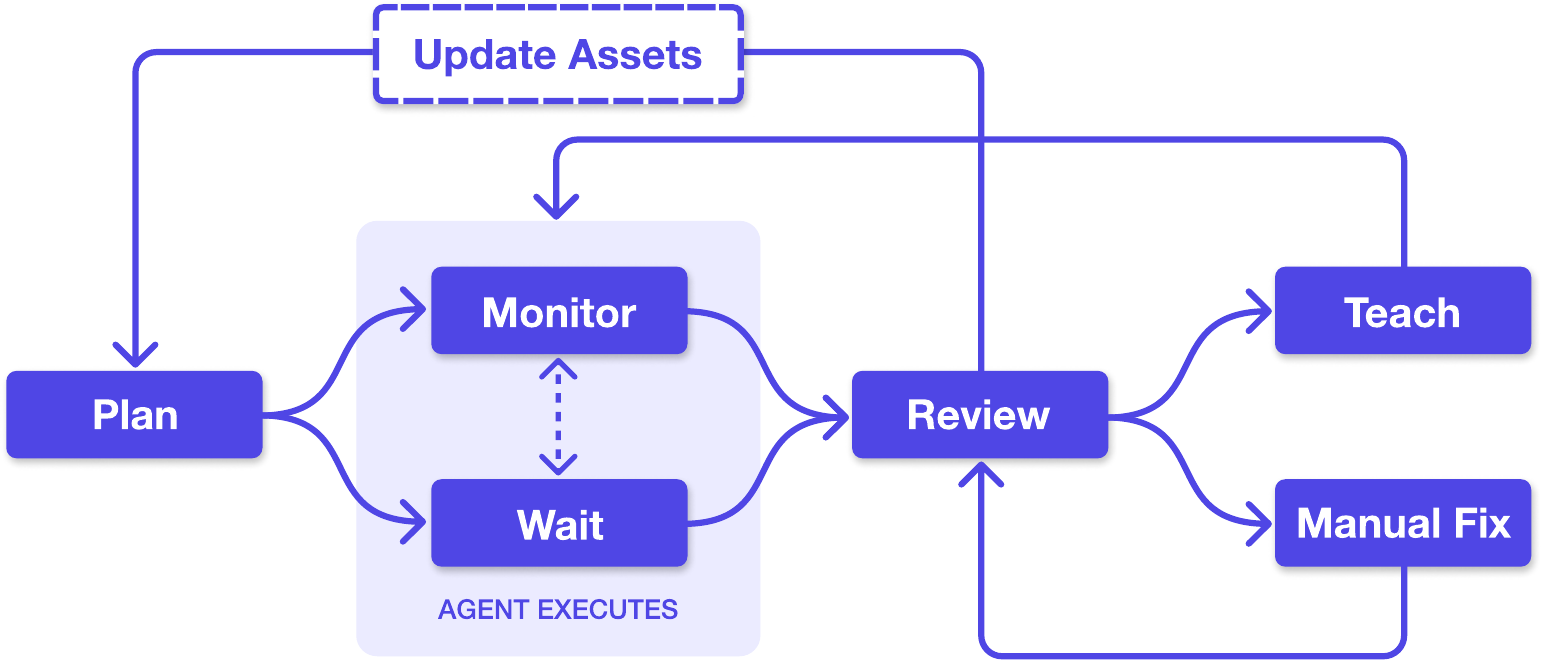}
    \caption{Our framework for supervising AI coding agents, consisting of seven stages—\textsc{Plan}, \textsc{Monitor}, \textsc{Wait}, \textsc{Review}, \textsc{Teach}, \textsc{Manual Fix}, and \textsc{Update Assets}—and the loops connecting them. \textsc{Monitor} and \textsc{Wait} occur while the agent executes, and developers can alternate between them. From \textsc{Review}, developers can go to either \textsc{Teach} or \textsc{Manual Fix}. \textsc{Teach} loops back to \textsc{Monitor}/\textsc{Wait}, while \textsc{Manual Fix} loops back to \textsc{Review}. \textsc{Update Assets} is an optional stage that feeds learnings from \textsc{Review} back into the next \textsc{Plan}.} 
    \label{fig:main}
    \Description{A flow diagram with a row of boxes---Plan, then a shaded region labeled "agent executes" containing Monitor and Wait side by side with a double-headed arrow between them, then Review, then Teach and Manual Fix side by side---connected left to right by arrows in sequence. An arrow loops from Teach back up into the Monitor/Wait region, and another arrow loops from Manual Fix back down into Review. Above the row, a dashed box labeled Update Assets is connected by two arrows: one arrives from Review, and one leaves toward Plan.}
\end{figure*}

We developed a new framework for supervising AI coding agents that reconfigures Sheridan's framework of human supervisory control based on our empirical findings. Our framework identifies seven distinct supervisory stages: \textsc{Plan}, \textsc{Monitor}, \textsc{Wait}, \textsc{Review}, \textsc{Teach}, \textsc{Manual Fix}, and \textsc{Update Assets}. These stages are connected through transitions and loops that capture the overall process of developer supervision across a development task, as illustrated in Figure~\ref{fig:main}. 

In this section, we introduce our framework by describing the stages in the general order they appeared in participants' workflows, grouping together those with similar supervisory goals. For each stage, we explain what developers did, how it relates to Sheridan's framework, and how it connects with other stages.

\vspace{0.25cm}
\vspace{-0.2cm}

\paragraph{\textbf{\textsc{Plan}}}
\textsc{Plan} corresponds closely to Sheridan's planning stage, in which the supervisor sets a goal and an overall approach before execution begins. At this stage, developers define the work to be delegated to the agent and establish its boundaries in advance, specifying goals, requirements, constraints, success criteria, and an intended approach. For example, P10 described, \textit{``I usually write three things [in the plan]: the situation, the success criteria, and the cautions. Here is the situation I am in, here is what you need to implement, and here is what you need to watch out for.''}

In Sheridan's framework, the supervisor determines the task and approach during \textsc{Plan}, then separately translates it into commands the system can execute during \textsc{Teach}. With AI coding agents, this translation was often unnecessary. Developers write the plan down in natural language with sufficient detail for the agent to act on directly. P3 explained, \textit{``I did planning, and the planning itself was the teaching. I wanted the teaching step to happen inside it.''} Hence, \textsc{Plan} serves as both the developer's specification and the agent's initial instruction, absorbing the initial \textsc{Teach} step.

In our framework, \textsc{Plan} also forms a self-loop. Developers refine their goals, requirements, and approach over multiple iterations before execution begins. These iterations include the developer's own revisions as well as input from the agent or another AI, which proposes, elaborates, or critiques the plan. P18 said, \textit{``There's a loop here, so if I don't like it I keep going around, revising [with AI] until I'm satisfied. And if the plan is right, then I assign the work.''} The loop ends when the plan aligns with the developer's intent and is detailed enough for the agent to act on directly. Developers then hand it over, and the workflow moves to \textsc{Monitor} or \textsc{Wait} as the agent begins executing.

\paragraph{\textbf{\textsc{Monitor} \& \textsc{Wait}}}
\textsc{Monitor} corresponds closely to Sheridan's monitoring stage, where supervisors observe an automated system during execution to track its progress and detect potential problems. Developers similarly observe agents as they work by reading their reasoning, watching the commands and file changes they produce, and checking intermediate outputs as they appear. P1 explained, \textit{``Definitely did the monitoring of the process to check if there are any easily identifiable errors happening in the reasoning or in the overall process.''} In our framework, \textsc{Monitor} forms a self-loop similar to Sheridan's monitoring loop. Developers continue observing the same ongoing execution and sometimes send additional prompts in response to what they see, without stopping the agent.

However, developers do not always monitor agents continuously during execution. When an agent can continue working for extended periods without human input, participants sometimes left it running and turned their attention elsewhere, describing this as a distinct \textsc{Wait} stage. They explicitly distinguished these periods from \textsc{Monitor}. During \textsc{Wait}, some developers shift to supervising agents on separate tasks, while others turn to unrelated activities. P2 explained, \textit{``I instruct from the planning stage, then wait, and while waiting I go on to another round of planning and instructing.''} On the other hand, P9 said, \textit{``I do something else---other work, or I just watch YouTube.''}

\textsc{Monitor} and \textsc{Wait} can alternate within the same execution as developers shift their attention between the agent and other activities. P3 described, \textit{``Sometimes I read them and sometimes I don't. I read for a bit, then go do something else, and then when I read and something seems really off, I stop it.''} When execution finishes, developers move to \textsc{Review}. When monitoring reveals a fundamental problem, they stop the execution and begin a new cycle from \textsc{Plan}.

\paragraph{\textbf{\textsc{Review}}}
\textsc{Review} is a new supervisory stage in which developers assess the agent's completed work against their intent and requirements. While Sheridan includes such assessment within \textsc{Monitor}, participants treated it as a separate \textsc{Review} stage. Developers inspect the code and diffs, run tests, or interact with the implemented feature to assess whether the changes behave as intended and satisfy their requirements. P18 explained, \textit{``[The plan] cannot capture all of my intent---some things get compressed, or it fills them in by guessing, and those only become visible once you get to this [Review] stage. [...] What I look for in [Review] is the gap between my intent and the result.''}

\textsc{Review} also forms a self-loop because developers rarely evaluate the agent's work in a single pass. Instead, they continue checking it in multiple ways until they are confident that it meets their expectations. For example, P9 described, \textit{``I first check how many of the tests pass---the acceptance conditions I gave the AI and the tests that match them. When those pass, I then check the feature with my own eyes, monkey-testing it.''} When developers are satisfied with the result, they complete the current task and move to \textsc{Plan} for a new task, sometimes passing through \textit{Update Assets} first. When they find errors or aspects they are dissatisfied with, they move to \textsc{Teach} or \textsc{Manual Fix}.

\paragraph{\textbf{\textsc{Teach} \& \textsc{Manual Fix}}}
When \textsc{Review} reveals errors or mismatches with the intent, developers supervise the correction in two ways. They either redirect the agent through \textsc{Teach} or take over the correction themselves through \textsc{Manual Fix}. In Sheridan's framework, \textsc{Teach} translates an intended action into commands that the automated system can execute, following either initial planning or corrective intervention. With coding agents, the initial teaching function is absorbed into \textsc{Plan}, leaving only corrective instructions as a distinct \textsc{Teach} stage (see Figure ~\ref{fig:teaser}).

\textsc{Teach} takes the form of follow-up prompting. Developers prompt the agent again based on problems identified during \textsc{Review}, adding missing details, correcting misunderstandings, reporting errors, or specifying how the work should change. P14 described this process, \textit{``Most of the time, it's because I didn't communicate something. [...] I realize, `I left this out,' and tell [the agent] again. [...] If there's actually a bug, I just tell it there's a bug, and it usually fixes it.''}

\textsc{Manual Fix} is where developers choose to take over and do the correction themselves. Developers make the correction when the change is simple or faster to carry out directly, or when they expect further prompting to be inefficient. P6 explained, \textit{``I could have had [the agent] do it, but it was faster for me, so I just changed it by myself.''}

After \textsc{Teach}, developers return to \textsc{Monitor} as the agent resumes execution, and then to \textsc{Review}, while \textsc{Manual Fix} directly leads back to \textsc{Review}. In both cases, the loop repeats until developers are satisfied with the result in \textsc{Review}.

\paragraph{\textbf{\textsc{Update Assets}}}
\textsc{Update Assets} is an optional stage that extends supervision beyond the current task by updating persistent resources that can guide future interactions with the agent. These resources include files such as \texttt{AGENT.md}, project documentation, and other shared assets that record project knowledge, preferences, constraints, or lessons from prior work. Developers do not necessarily update these assets after every task, but when they do, the updated assets support future tasks, beginning with a new \textsc{Plan} stage. In this way, Sheridan's \textsc{Learn}, which focuses on updating the supervisor's understanding from experience, becomes \textsc{Update Assets} as developers externalize what they have learned into persistent resources that can guide the agent in future work. P4 explained, \textit{``[Learning] also happens partly at the system level, because the company's skills, assets, and data accumulate. [...] Those assets then feed back into planning.''}

%% file: sections/6_Application.tex
\section{Applying the Framework to Developer Discussions}

Using our framework as an analytical lens, we examined public developer discussions to identify supervisory patterns that extend beyond individual instances. The framework allowed us to bring together practices across discussions and examine recurring patterns across the supervision process. Specifically, we focused on (1) what required developer involvement as more execution was delegated to agents and (2) how developers adapted their supervision in response.

We chose Reddit\footnote{\url{https://www.reddit.com/}} as a source of developer discussions because its developer communities host naturally occurring, in-depth discussions of concrete experiences with AI coding agents~\cite{klemmer2024using, pimenova2025good, baltes2026endless}. These discussions span diverse tasks, codebases, and agent setups, providing large-scale evidence across development contexts. Hence, we collected discussions from three developer subreddits and used our framework to systematically identify and analyze supervisory practices.

\subsection{Data Collection} \label{sec:reddit_collection}
We collected Reddit discussions and screened them to obtain a final sample of 102 threads with 12,912 comments. We describe this data collection and screening process in this section.

We selected three developer subreddits for data collection: \texttt{r/ClaudeCode}\footnote{\url{https://www.reddit.com/r/ClaudeCode/}}, \texttt{r/codex}\footnote{\url{https://www.reddit.com/r/codex/}}, and \texttt{r/ExperiencedDevs}\footnote{\url{https://www.reddit.com/r/ExperiencedDevs/}}. We chose \texttt{r/ClaudeCode} and \texttt{r/codex} because they are active communities centered on two widely used coding agents, and \texttt{r/ExperiencedDevs} because it restricts participation to developers with at least three years of experience. Together, these communities capture both agent-focused discussions and perspectives from experienced developers.

Given the rapid evolution of agentic tools and practices, we focused on a recent three-month period (May 17--August 17, 2026) to capture current practices at sufficient scale. Using Project Arctic Shift~\cite{arcticshift}, we first collected the posts from the three subreddits during this period, including each post's title, body, timestamp, and other metadata. This yielded 41,757 posts in total: 23,060 from \texttt{r/ClaudeCode}, 16,271 from \texttt{r/codex}, and 2,426 from \texttt{r/ExperiencedDevs}.

We next screened these posts to identify threads with substantial evidence of supervisory practices for qualitative analysis. We first removed posts that did not provide useful data: (1) posts with deleted content, (2) posts with no comments, and (3) posts with less relevant flairs (e.g., \textit{Humor}, \textit{News}; see Appendix~\ref{app:reddit_filters} for details), leaving 11,471 posts. 

We then used \texttt{gpt-5.6-luna}\footnote{\url{https://developers.openai.com/api/docs/models/gpt-5.6-luna}} in two screening steps to further narrow the corpus. In the first step, the model used each post's title and body to conservatively exclude posts unrelated to the use of AI in software development, retaining 5,041 posts. For these posts, we collected all comments to reconstruct the complete threads. In the second step, the model tagged verbatim excerpts corresponding to any of the seven supervision stages defined in Study 1, identifying 4,390 threads with at least one excerpt. The prompts used for both screening steps are provided in Appendix~\ref{app:llm_prompts}.

We ranked these threads by the number of tagged excerpts and selected the top 100, including ties, for a total of 103 threads. We then manually checked each thread for relevance, excluding one that primarily promoted a product. The final sample consisted of 102 threads with 12,912 comments: 70 from \texttt{r/ClaudeCode} (9,151 comments), 26 from \texttt{r/codex} (2,631 comments), and 6 from \texttt{r/ExperiencedDevs} (1,130 comments).

\subsection{Thematic Analysis}
We conducted an inductive thematic analysis of the selected 102 threads following the method of ~\citet{braun2006thematic}. We began by randomly sampling 30 threads (29.4\%) to first develop an initial codebook. Three authors first divided these sample threads and read each discussion thoroughly, tagging all relevant verbatim excerpts for the seven supervision stages. The prior LLM tagging results were visible during this pass, and we revised or removed them where they did not match our own judgment. This produced 1,114 evidence quotes: 366 for \textsc{Review}, 230 for \textsc{Update Assets}, 204 for \textsc{Plan}, 127 for \textsc{Teach}, 74 for \textsc{Wait}, 73 for \textsc{Monitor}, and 40 for \textsc{Manual Fix}.

We pooled the excerpts by each stage to preserve their context and familiarized ourselves with the data. The three authors then independently open-coded the excerpts within each stage, focusing on supervisory demands and the strategies developers used to address them. We compared our codes, merged convergent ones, discussed differences in interpretations, and consolidated them into an initial codebook. Two authors applied the initial codebook to the remaining 72 threads, discussing new or ambiguous cases and iteratively revising the codebook. One additional code emerged during this process, and no further changes were needed. Finally, we compared the resulting codes across stages and grouped them into themes that captured patterns in demands and strategies across the supervision process.

%% file: sections/7_Findings.tex
\section{Findings}

Applying our framework to developer discussions revealed patterns across the supervision process. We present our findings on (1) the supervisory demands developers encountered and (2) the strategies they used to address them.

\subsection{Demands of Supervising AI Coding Agents}
Across the seven supervisory stages, we identified three demands that spanned multiple stages: keeping agent work aligned with developer intent, enforcing quality standards beyond task completion, and maintaining understanding of agent-produced work. Each demand appeared at different points in the workflow, taking different forms as developers planned, monitored, reviewed, and corrected agent work. We describe these three demands below.

\subsubsection{Keeping Agent Work Aligned with Developer Intent} \label{sec:demand_intent}
A key supervisory demand was ensuring that agent work remained aligned with the developer's intent throughout the workflow. During \textsc{Plan}, developers established the goals, requirements, and boundaries that the agent was expected to follow, but agents could interpret the plan differently, change its criteria, make unintended assumptions, or expand the task scope on their own.

During \textsc{Monitor}, developers had to detect when the agent's work began to diverge from the intended plan. This divergence took several forms. Agents sometimes changed the plan itself, as one developer emphasized the need to ensure that it \textit{``doesn't make up criteria or change criteria as it goes.''} Others skimmed the agent's conversation to \textit{``spot how it deviates in an unwanted direction.''} Execution could also stall rather than progress as planned, leading one developer to set up \textit{``monitors for when my agents get hung or stuck in place.''} Developers therefore monitored the agent during execution to catch these problems before they accumulated into larger amounts of misaligned work.

The same demand extended into \textsc{Review}, where developers assessed alignment again after the agent had completed its work. Here, they checked both whether the implementation had become unnecessarily complex and whether the agent had introduced work beyond the intended scope. One developer described how review \textit{``checks both things: did we make it more complicated than needed, and did the agent actually build what was in the plan instead of sneaking in extra work. If it did something outside the plan, that is a problem, even if the extra thing looks reasonable on its own.''}

When \textsc{Review} revealed misalignment, this demand carried into \textsc{Teach}, where developers redirected the agent toward the original plan. One developer explained, \textit{``If I catch it over-engineering, I can give some push back and it makes the appropriate change.''} However, repeated correction could itself produce drift. One developer described \textit{``scope drift over long sessions,''} where the agent \textit{``starts building on its own previous assumptions''} across successive fixes.

\subsubsection{Enforcing Quality Standards Beyond Task Completion} \label{sec:demand_quality}
Even when agents followed the intended plan, developers still faced a demand to ensure that the resulting code met their quality standards. Completing the requested functionality was not sufficient, as working code could still fall short in how it was written. One developer noted that agents often produced \textit{``lazy, sloppy, repetitive, and overly defensive code---even with solid guardrails and tests.''}

During \textsc{Review}, developers checked aspects of code quality beyond functional correctness and carried identified issues into \textsc{Teach} or \textsc{Manual Fix} for revision. One developer explained that \textit{``part of the validation phase should include duplicate/dead code checks, among many more review items.''} When such issues were found, developers could prompt agents to improve the implementation, as one developer described, \textit{``It really helps steering the next prompt to keep the code DRY and modular.''} Others manually made quality improvements themselves. For example, one developer \textit{``re-write[s] most of the important comments [...] to make sure it's easy to understand for other human reviewers.''}

\subsubsection{Maintaining Understanding of Agent-Produced Work} \label{sec:demand_learn}
As agents produced larger portions of the implementation, developers faced a supervisory demand to maintain enough understanding of work they had not written themselves. This involved keeping track of what the agent was doing, what had changed, and how the codebase was evolving.

During \textsc{Monitor} and \textsc{Review}, developers maintained understanding by following the agent's work and reviewing its results. One developer explained, \textit{``I like to read the code and talk with the model while it's working. That way I understand what's happening and stay engaged with it.''} Another described reading the code after execution to keep their \textit{``mental model of how the code works''}, calling it a way to \textit{``stay in control and to keep understanding the system that is being built.''}

The need to maintain understanding also appeared in \textsc{Manual Fix}. Developers sometimes remained directly involved in technically important parts of the implementation, particularly where broader system implications had to be understood. One developer explained, \textit{``I still code the tricky architectural decisions myself, the stuff where you need to think through the full system implications. [...] [It] forces you to understand what the agent is actually building.''}

\subsection{Strategies for Supervising AI Coding Agents}

We identified three strategies through which developers organized supervisory effort across the workflow: concentrating effort in planning, distributing supervisory work between developers and agents, and externalizing supervision into reusable assets. Together, these strategies shaped where supervisory effort was concentrated, who performed it, and whether it needed to be repeated in later work. We present these strategies below.

\subsubsection{Concentrating Supervisory Effort in Planning} \label{sec:strategy_plan}
Developers concentrated supervisory effort in the \textsc{Plan} stage, investing more effort upfront to reduce the supervision required later in the workflow. Developers described well-structured plans as allowing them to shift from \textsc{Monitor} to \textsc{Wait}, enabling them to have agents run \textit{``for days without interruption and come back to clean and efficient implementation of code''} and gain \textit{``free time to manage other agents.''} They also associated stronger plans with fewer repeated \textsc{Teach} cycles, as one developer noted that with a sufficiently tight architecture and testing plan, the \textit{``whole thing can run in literally one pass, no looping needed.''} Planning also helped developers build their own understanding before implementation was delegated, with one developer noting that they got \textit{``a very clear understanding of how the features should work, how the architecture should look, and which technical decisions needed to be made. A lot of problems were solved before any code was written.''} Below, we describe the strategies developers used when constructing plans, deciding what to include in them, and delivering those plans to agents.

\paragraph{\textbf{Construction of the Plan.}}
Developers often had a clear idea of what they wanted before planning, but their intent still had to be articulated with natural language in a form the agent could follow. Since the resulting plan served directly as the agent's instruction, developers involved agents in constructing it to expose assumptions, ambiguities, or missing details before execution. One developer brought a plan they had already drafted to the agent for stress testing, explaining that they would \textit{``sit down and plan it out. Architect the solution to a problem. Then I put it through the wringer with AI, stress testing it and looking to see if I missed anything. Then the plan gets drafted.''} Others used dedicated planning skills that prompted the agent to interview them for missing details. One developer would \textit{``describe feature/task and ask it to /grill-me on it,''} then \textit{``answer questions until the skill reaches full understanding.''}

Developers also constructed plans through multiple agents, using disagreement between them to expose gaps or weak assumptions in the plan. One developer asked two models to write plans from similar prompts and then had them \textit{``incorporate from each other's good ideas,''} letting \textit{``the one that did better consolidate and merge and then again the other to check it.''} Others separated plan generation from plan review. One developer had Claude prepare the plan while \textit{``Codex does the review, and critique is messaged back to Claude,''} after which \textit{``they loop till agreement.''}

\paragraph{\textbf{Components of the Plan.}}
Once the direction was settled, developers specified it through concrete constraints and representations that agents could act on. One set of components defined the boundaries of the work. Developers stated non-goals alongside goals to clarify the intended scope of a feature. One developer emphasized being \textit{``very explicit''} about these boundaries, giving examples such as \textit{``do not start x,'' ``do not rewrite y,''} and \textit{``preserve z.''} Developers also specified which parts of the codebase the agent was allowed to modify. One described defining a \textit{``change contract''} before prompting the agent, including \textit{``allowed files, expected behavior, and what it must not touch.''}

Another set of components represented the intended solution more concretely. Some developers provided examples of the desired result, with one recommending that, for a new feature, developers first write \textit{``the example you \emph{wish} worked''} and then have the agent build toward it. Others supplied architecture diagrams or pseudocode to specify the intended structure. One developer described giving the model their own pseudocode to fill in, explaining that this \textit{``ensures that its only building what I ask, and also importantly ensures that I know what the design of the codebase is.''}

\paragraph{\textbf{Delivery of the Plan to Agents.}}
Developers structured how plans were delivered to agents because, as one developer put it, agents were \textit{``great at executing, but pretty bad at preserving product direction over time''} unless developers imposed structure around them. One strategy was to divide planned work into smaller execution units. One developer described breaking features down \textit{``more than I would have done when I was manually doing it,''} reporting that this brought \textit{``a lot more success [...] [and] seems to produce higher quality.''} Developers then paired these smaller units with explicit end conditions. One developer explained, \textit{``Write the definition of done before it starts. One line: `done = this test passes and nothing outside file X changes.' Then anything it wants to do that isn't that is, by definition, a rabbit hole.''}

Developers also kept the plan visible throughout execution instead of relying on agents to retain it from the initial instruction. One developer required \textit{``a detailed architecture plan''} and \textit{``refer[red] to that plan in each step''}  as implementation progressed. For longer work, developers managed context so the plan remained salient, for example by using separate branches to keep \textit{``the context window clean enough that it doesn't drift''} or external tools that \textit{``re-pin your goal + context health to the bottom of the window every prompt so the model doesn't lose the plot.''}

\subsubsection{Distributing Supervisory Work} \label{sec:strategy_distribute}
Developers addressed supervisory demands by distributing supervisory work between themselves and agents. They decided which supervisory tasks agents could perform independently, which required developer involvement, and which developers should retain entirely, covering the workflow without handling every demand directly. We describe these three arrangements below.

\paragraph{\textbf{Delegated Supervision.}}
Developers delegated parts of supervisory work to agents or automated mechanisms, allowing some supervisory loops to proceed without direct developer involvement. Such delegation required deciding in advance what would be watched, how detected problems should be handled, and when the corrective cycle should stop. Once configured, these arrangements allowed supervisory loops to proceed without developer involvement. In \textsc{Monitor}, one developer assigned other agents to detect when an agent became \textit{``hung or stuck in place,''} so that they could \textit{``help the original agent out of the hole.''} In \textsc{Teach}, developers set up feedback loops in which one agent reviewed another agent's implementation and passed problems back for correction, as in an arrangement where an \textit{``Advisor checks the implementation, calls out anything bad, back to Implementer, repeat till it's clean.''}

\textsc{Review} was delegated to agents most extensively, and developers were deliberate about how they configured the reviewing agents. One approach was to separate the reviewer from the implementer by model provider. One developer used \textit{``multiple models from different providers to have them all audit each other's work,''} explaining that each model had \textit{``its own biases and blindspots''} and that \textit{``getting a second opinion from a different architecture is extremely useful.''} Another approach was to separate reviews by focus, running \textit{``one pass for security/data loss, one for test gaps, one for overengineering,''} because \textit{``when one review tries to cover everything, it tends to produce generic advice.''}

Developers also used hooks to make parts of these supervisory loops run automatically and reliably. One developer contrasted hooks with written guidance, which could be \textit{``read, weighed against everything else, and quietly deprioritised once context fills up,''} whereas \textit{``a hook is code; it fires every time whether the model thinks it's relevant or not.''} This allowed developers to embed specific supervisory actions directly into the workflow. For example, one used a stop hook that prompted an agent entering a particular area of the codebase to \textit{``first read the system patterns for that area.''} Another used a PostToolUse hook that ran \textit{``the narrowest relevant test right after an edit''} and fed failures back immediately, so the agent learned it had broken something \textit{``on the same turn, before it stacks three more edits on a broken base.''}

\paragraph{\textbf{Human--Agent Collaborative Supervision.}}
Developers also shared supervisory work with agents, allowing agents to handle ongoing checks while retaining points for human inspection and judgment. In \textsc{Monitor}, agents handled routine coordination, while developers were brought in when needed. One developer built a relay that \textit{``passes messages between Claude and Codex,''} but \textit{``if anything weird happens it always default freezes the relay and alerts me.''}

This collaboration was especially common in \textsc{Review}. Developers reviewed the work after multiple rounds of agent review, saying \textit{``agents and humans catch different types of errors.''} They also divided review so agents handled more routine checks, letting developers focus on what remained. As one explained, \textit{``by the time I look at a change I'm reviewing something that's already been checked for obvious mistakes, failed validations, and deviations from the task specification.''}

\paragraph{\textbf{Developer-Retained Supervision.}}
Developers sometimes retained supervisory work because delegating it to the agent carried its own cost. For small changes, explaining and routing the work through an agent could require more effort than making the change directly. One developer therefore made \textit{``surgical edits for things that would take longer to type into a prompt than to just fix myself.''} Repeated correction created a similar cost, with one finding it \textit{``mentally much less taxing to just code it up manually''} than to keep prompting the agent for the same outcome.

Developers also retained supervision to maintain their own understanding of the work. One developer described they \textit{``read the reasoning to steer it when needed or learn more about how to properly direct it the next time I prompt.''} Another periodically rewrote \textit{``one risky function or migration without the agent, just to keep the map in my head.''}

\subsubsection{Externalizing Repeated Supervision into Assets} \label{sec:strategy_assets}
Developers also externalized parts of supervision into reusable assets rather than repeating the same work in each interaction. Some assets guided agents directly, while others helped developers recover context and maintain understanding. Although creating and maintaining these assets required effort, they reduced supervisory demands that would otherwise recur later. We describe each below.

\paragraph{\textbf{Assets for Agents.}}
Developers turned recurring supervision into assets that reduced the need to repeat the same intervention in later work. As one developer put it, \textit{``if I catch myself repeating the same instruction to the agent, that's not a prompting problem, that's a missing hook.''} One developer, after getting \textit{``tired of writing the same review comments,''} built \textit{``a static linter that checks for this automatically before commit.''} Another maintained a \textit{``MISTAKES.md file where every mistake the agent makes gets documented, along with a rule in CLAUDE.md telling it to record them.''} Developers also used completed work to strengthen these assets for future supervision. One ran a retrospective at the end of each milestone, where the \textit{``retro edits or writes skill files and meta-learnings as memory.''} Across projects, this produced \textit{``compounding effects, essentially preventing the same kind of mistakes from happening in the future.''}

\paragraph{\textbf{Assets for Developers.}}
Developers used assets to reduce the effort required to maintain their own understanding of agent-produced work. Instead of reconstructing the purpose, decisions, and current state of the work each time, they kept records they could return to later. One developer recommended a feature document describing \textit{``what this code is supposed to do without any implementation details,''} explaining that this \textit{``helps agents and humans track the context of this work.''} Others used assets to recover their context after stepping away. One described a \textit{``wrap-up skill that takes what your are doing and next steps and puts it in a file,''} so that on returning, it could \textit{``remind you what's going on.''}

%% file: sections/8_Discussion.tex
\section{Discussion}

In this work, we developed a framework for supervising AI coding agents and applied it to developer discussions. Our findings suggest that delegating execution to agents redistributes supervisory work across the development process. Building on these findings, we discuss the framework's applicability beyond software development, design implications for agentic systems, and how the framework can be further extended as agent autonomy increases.

\subsection{Applicability of the Framework}
Our framework captures supervisory functions that may extend beyond software development to other forms of delegated agentic work. Software development has been at the forefront of agentic delegation. Similar forms of delegation have now started to emerge in knowledge and creative work, where agents support tasks such as writing, research, data analysis, and design~\cite{googlecloud2026geminiagentplatform, openai2026chatgptwork, figma2026agent}. A particularly relevant example is computer-use agents, which can carry out tasks across browsers, documents, spreadsheets, and communication tools with limited human involvement~\cite{openai2025operator, anthropic2026fable51, openai2026gpt6astra}. As users delegate such work, they may need to specify goals and constraints, oversee execution, evaluate outcomes, and intervene when necessary. Thus, our framework may generalize beyond software development, providing a common structure for analyzing these supervisory functions across domains.

Applying the framework in these settings would require adapting each supervisory stage to the characteristics of the delegated work. In many knowledge and creative tasks, goals and quality criteria may be less precisely specified, and outputs cannot always be verified through deterministic checks. For example, in design tasks, \textsc{Plan} may involve specifying target users, interaction goals, and visual constraints, while \textsc{Review} may rely more on human judgment of usability, coherence, and aesthetic quality. Applying the framework across such domains could reveal how supervisory effort is distributed differently across types of work and where new demands for human involvement arise.

\subsection{Design Implications for Agentic Systems} 

We present design implications for supporting developer supervision of AI coding agents.

\subsubsection{Developing AI Coding Agents}
Our findings suggest several directions for developing AI coding agents that require less supervisory effort from developers. First, developers of AI coding agents should prioritize preserving developer intent throughout execution. Developers faced supervisory demands when agents deviated from intended goals, constraints, or scope (\S\ref{sec:demand_intent}). Thus, agents should be designed to consistently follow the plans and constraints provided by developers, reducing the need for repeated monitoring and redirection. For example, methods for maintaining goals and constraints over longer interactions~\cite{liu-etal-2026-context} or detecting when an agent's planned actions conflict with previously specified intent~\cite{BatoleKhomhRajan2026refineact} could help prevent such deviations before developer intervention becomes necessary.

Second, developers of AI coding agents should account for quality expectations beyond functional correctness. Even when functionality was correct, developers had to address implementations they considered repetitive, overly defensive, or difficult to maintain (\S\ref{sec:demand_quality}). Hence, agents should be designed to reflect developer- and project-specific preferences in how code is produced. For instance, preference-alignment methods~\cite{sivapiran2026reward,rafailov2023direct} could be used to align agent outputs with preferred coding conventions, reducing the need to repeatedly correct the same quality issues.

\subsubsection{Supporting Developer Supervision}
Our findings also suggest several ways systems can be designed to better support developers in supervising AI coding agents. First, systems can better support developers in the planning stage prior to task delegation. Developers invested substantial effort in constructing plans that clarified goals, boundaries, and success criteria and reduced supervision later in the workflow (\S\ref{sec:strategy_plan}). Systems could help developers articulate these elements, identify missing or ambiguous requirements, and keep plans accessible throughout execution~\cite{feng2026cocoa, vaithilingam2025sematiccommit, kim2024evallm}.

Second, systems can be designed to reduce the effort of setting up and maintaining delegated supervision. Developers reduced supervisory work by delegating \textsc{monitor} and \textsc{review} to other agents (\S\ref{sec:strategy_distribute}) and using reusable assets such as instruction files, tests, and hooks for repeated supervision (\S\ref{sec:strategy_assets}). However, these arrangements required developers to configure and maintain multiple agents and assets. Systems could help developers manage these forms of delegated supervision together, making it easier to inspect, update, and reuse how supervision is set up for a task~\cite{liu2026masfactory, tan2026skillcoverage, pan2026nlagentharnesses}.

Finally, systems can be designed to help developers maintain an understanding of work delegated to agents. Developers actively followed and revisited agent-generated code to maintain a mental model of the codebase over time (\S\ref{sec:demand_learn}). Systems could support this by highlighting important changes, connecting them to the agent's rationale, and summarizing the context needed for developers to understand what changed and why~\cite{yan2024ivie, kazemitabaar2024improving, hutter2026agentstepper}.

\subsection{Further Development of the Framework}
Although our framework captures a broad range of supervisory activities in agentic software development, the structure of supervision may change as agent autonomy increases. Prior work suggests that as AI agents become more autonomous, users may shift from actively directing their work toward approving or even observing it~\cite{feng2025levels, zhou2026should, choudhuri2026autonomy}. Such changes may redistribute human involvement across the supervisory process, making some stages less prominent while increasing the importance of others. For example, practitioners increasingly debate whether every agent-generated change requires human review, with some arguing for human review only for high-risk or consequential changes~\cite{laycock2026codereview, shukla2026hedwig}. Hence, the framework can be further developed to reflect these changes as new forms of human--agent supervision emerge.

The framework could also be developed to incorporate evidence about the effectiveness of different supervisory practices. Empirical studies could examine how the allocation of supervisory effort across stages relates to development outcomes. For example, they could test whether greater effort in \textsc{Plan} reduces later correction, delegated \textsc{Review} improves code quality, or different supervisory workflows affect developer time and errors~\cite{lai2022human}. Such evidence could add an evaluative layer to the framework, indicating which forms of supervision are more effective under what conditions.

Finally, the framework could be extended to incorporate the effects of supervision on developers. As implementation is increasingly delegated to agents, different supervisory arrangements may shape developers' agency~\cite{pu2025assistordisrupt, schlonsak2026costofconvenience}, reliance on AI~\cite{kim2025fostering, ibrahim2026overreliance, he-etal-2026-recap}, self-efficacy and authorship~\cite{park2026authorship, seo2026whosecode}, and ability to maintain skills and understanding over time~\cite{qianou2026pedagogical, mitchell2026aiagentspushhumans}. The framework could be extended to connect supervisory arrangements with their longer-term effects on developers.

%% file: sections/9_Limitation.tex
\section{Limitations and Future Work}

We acknowledge several limitations of our study and suggest potential future work.

First, our study for framework development captured only a limited window of participants' development work. Although we asked participants to bring tasks that would allow us to observe an end-to-end supervisory workflow within 40 minutes, real-world development projects often span longer periods and multiple sessions beyond what we could capture. The subsequent workflow design activity helped participants extend their diagrams beyond the observed task by considering other tasks and developers, but this still relied on retrospective reflection. Future work could examine longer-term supervision through longitudinal observations~\cite{sergeyuk2026evolving, he2026cursor, vella2026impactaicodingassistants} or interaction traces~\cite{tang2026programming}.

Second, participants' observed behavior and resulting workflow diagrams may have been affected by the study setting. Participants completed their tasks while being observed and thinking aloud, which may have altered how they worked. Moreover, although we introduced Sheridan's framework to help participants reflect on how their workflows related to it and emphasized that they need not follow its structure, prior exposure may still have anchored how they segmented or represented their supervision. Future work could complement our approach with less intrusive observations and workflow elicitation without prior exposure to a theoretical framework.

%% file: sections/10_Conclusion.tex
\section{Conclusion}

In this work, we developed a framework for supervising AI coding agents, reconfiguring Sheridan's framework of \textit{human supervisory control}. We then applied our framework to developer discussions on agentic AI in software development and found that supervisory demands extend across stages. Developers manage such demands by shifting effort across the workflow, delegating parts of supervision, and creating reusable assets. These findings highlight that increasing autonomy of agents changes where and how human involvement is needed. Our framework provides a lens for understanding and supporting human supervision as agentic software development continues to evolve.

%% file: sections/99_Appendix.tex
\section{Pre-Task Interview Questions} \label{app:interview_questions}

We list the full interview questions for the study below.

\subsection{Participant Background}

\begin{itemize}
    \item Could you briefly introduce yourself and your current role?
    \item How much experience do you have with software development or programming?
    \item When did you start using AI coding agents?
    \item Could you briefly describe the AI coding tools or agent setup you usually use?
    \item When you delegate work to AI coding agents, what level of granularity do you usually use?
    \item How do you usually review or verify the outputs generated by AI coding agents?
    \item In your current company, team, lab, open-source project, or other work context, are there any explicit or implicit rules, norms, or expectations around using AI coding tools? 
    \item If so, how do they affect the way you use these tools in practice?
    \item Compared to your colleagues or peers, in what ways do you feel your use of AI coding tools is similar or different?
\end{itemize}

\subsection{Prepared Task}

\begin{itemize}
    \item Could you briefly describe the task you prepared for today?
    \item What would you like to accomplish within the 40-minute coding session?
\end{itemize}

\clearpage
\section{Example Supervision Workflow Diagrams} \label{app:workflow_diagrams}
We present example supervision workflow diagrams drawn by participants in our interviews, illustrating the diversity of stages and loop structures observed. The diagrams shown here are translated to English from the original version.

\begin{figure}[H]
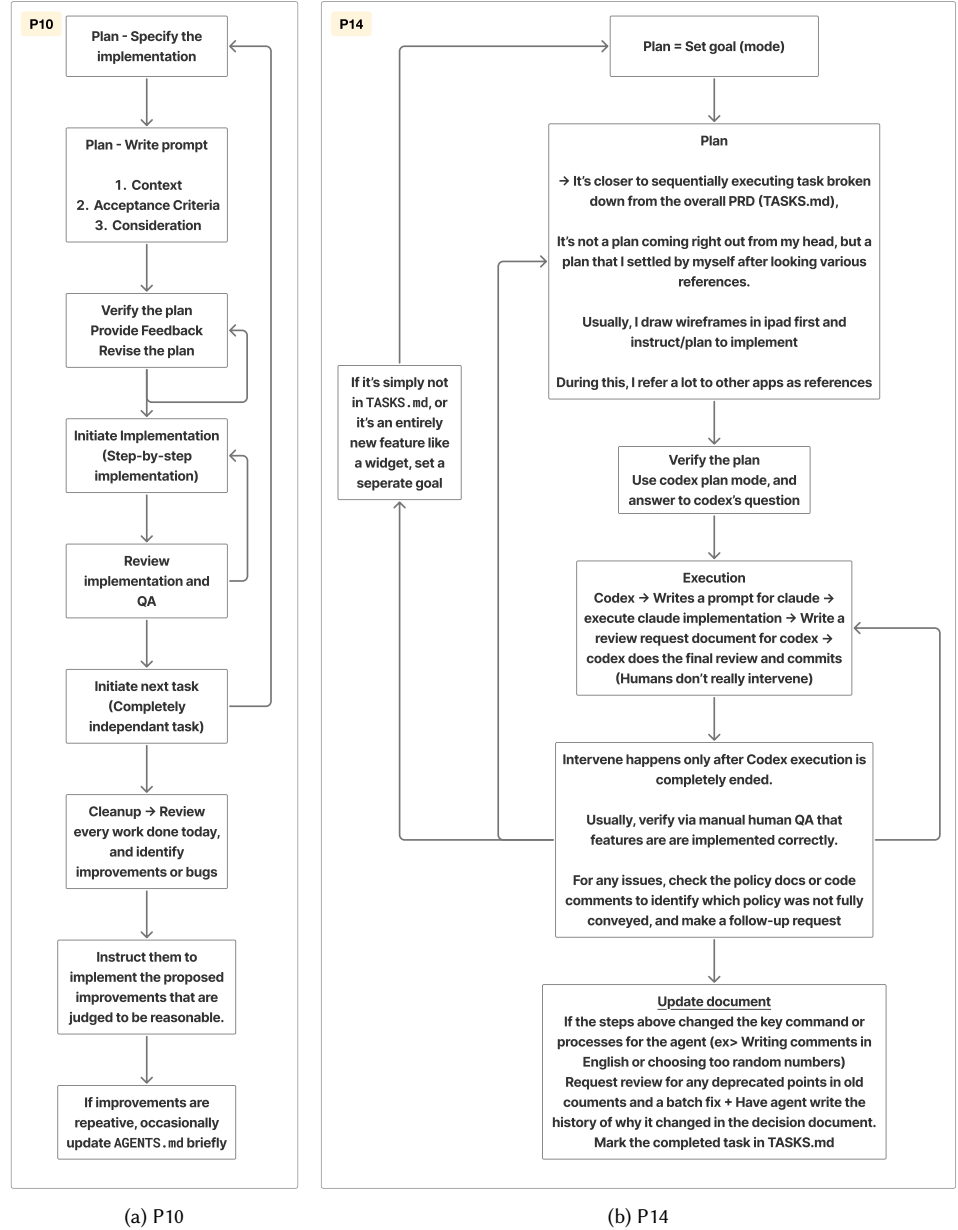

  \centering
  \begin{minipage}{0.635\textheight}
    \centering
    \begin{subfigure}[t]{0.30401456\linewidth}
      \centering
      \vspace{0pt}
      \includegraphics[width=\linewidth]{figures/P10.pdf}
      \caption{P10}\label{fig:sample_diagram_p10}
      \Description{Flow diagram of P10's supervision workflow: a vertical sequence of stages with feedback loops.}
    \end{subfigure}%
    \hfill
    \begin{subfigure}[t]{0.67198544\linewidth}
      \centering
      \vspace{0pt}
      \includegraphics[width=\linewidth]{figures/P14.pdf}
      \caption{P14}\label{fig:sample_diagram_p14}
      \Description{Flow diagram of P14's supervision workflow: a branching sequence with feedback loops connecting back to earlier stages.}
    \end{subfigure}%
  \end{minipage}
  \caption{Example supervision workflow diagrams from participant interviews, showing diverse stages and loop structures.}
  \Description{Two participant-drawn supervision workflow diagrams shown side by side: P10 (narrower, left) is a linear vertical sequence of stages with several feedback loops; P14 (wider, right) is a more branching diagram with parallel goal-setting paths and a distinct post-execution intervention stage.}
  \label{fig:sample_diagrams}
\end{figure}
\clearpage

\section{Reddit Flair Filtering} \label{app:reddit_filters}

On Reddit, post \textit{flairs} are labels used to categorize posts within a subreddit. We used these flairs to exclude posts unlikely to contain relevant discussions before LLM-based screening. Since each subreddit uses its own set of flairs, we applied subreddit-specific filtering criteria. The flairs excluded for each subreddit are listed below.

\paragraph{\texttt{r/ClaudeCode}.}
We retained posts with the following flairs: \textit{Discussion}, \textit{Question}, \textit{Help Needed}, \textit{Help/Question}, \textit{Resource}, \textit{Tutorial / Guide}, and \textit{Tips \& Workflows}. On the other hand, we excluded posts with the following flairs: \textit{Showcase}, \textit{Humor}, \textit{Bug Report}, \textit{Built with Claude}, \textit{Meta}, \textit{Solved}, \textit{Bug / Issue}, \textit{Rant}, or \textit{News/Updates}, as well as variants of \textit{Megathread}, \textit{Weekly Showcase}, and \textit{Community Update}. This filtering step retained 7,232 of 10,427 posts.

\paragraph{\texttt{r/codex}.}
We retained posts with the following flairs: \textit{Question}, \textit{Complaint}, \textit{Commentary}, \textit{Suggestion}, \textit{Workaround}, or \textit{Instruction}. On the other hand, we excluded posts with the following flairs: \textit{Showcase}, \textit{Bug}, \textit{Praise}, \textit{Limits}, \textit{News}, \textit{Other}, \textit{Comparison}, \textit{Humor}, \textit{Reset}, \textit{Megathread}, or \textit{MOD}. This filtering step retained 4,054 of 7,284 posts.

\paragraph{\texttt{r/ExperiencedDevs}.}
We retained posts with the following flairs: \textit{Career/Workplace}, \textit{AI/LLM}, and \textit{Technical question}. We excluded posts labeled \textit{Meta} or \textit{Big Tech}, as well as posts without a flair. This filtering step retained 185 of 211 posts.

\section{LLM Prompts Used} \label{app:llm_prompts}

We present the prompts used for the two LLM-assisted screening steps in our data collection of Reddit threads as explained in \S\ref{sec:reddit_collection}: (1) screening posts for relevance and (2) initial supervision stage tagging.

\subsection{Prompt for Screening Posts for Relevance}

\vspace{0.1cm}

\begin{tcolorbox}[
breakable, enhanced, top=1pt, left=1pt, right=1pt, bottom=1pt, colback=white, fontupper=\ttfamily, fonttitle=\bfseries\small,
title={Prompt}
]
\small
You are screening posts for a study of how developers supervise AI coding tools and agents during software development. \\
\\
Keep a post if it meaningfully discusses how developers work with, manage, or respond to AI in software-development work. This includes concrete practices, experiences, problems, failures, decisions, role changes, or advice related to AI-assisted development. \\
\\ 
Exclude posts that are only about: \\
\\
- Model releases, benchmarks, or general performance comparisons \\
- Pricing, subscriptions, tokens, rate limits, or usage quotas \\
- Product availability, outages, account issues, or other service-related problems \\
- General AI, career, or future-of-programming discussion with no concrete connection to AI-assisted development work \\
- Building, announcing, or promoting a tool or AI system without substantively describing a relevant development practice, workflow, problem, or experience \\
- Choosing between tools, plans, or editors without connecting the choice to concrete development tasks, workflows, or ways of working \\
- Memes, jokes, advertisements, or unrelated technical support \\
\\
A post may still qualify if it promotes or introduces a tool, as long as it substantively describes a relevant AI-assisted development practice, workflow, problem, or experience. \\
\\
A title-only or very brief post should be kept only when the available text itself clearly describes a relevant practice, problem, or experience. \\
\\
A post does not need to describe a complete workflow. A concrete discussion of even one relevant practice, experience, problem, failure, decision, or role change is sufficient. \\
\\
When uncertain, keep the post. \\
\\
Return only JSON: \\
\\
\{ \\
  ``keep'': true, \\
  ``confidence'': ``high | mid | low'', \\
  ``reason'': ``Brief reason''\\
\} \\
\\
Title:\\
\{\{TITLE\}\} \\
\\
Body: \\
\{\{BODY\}\} \\
\\
\end{tcolorbox}

\subsection{Prompt for Initial Supervision Stage Tagging}

\vspace{0.1cm}

\begin{tcolorbox}[
breakable, enhanced, top=1pt, left=1pt, right=1pt, bottom=1pt, colback=white, fontupper=\ttfamily, fonttitle=\bfseries\small,
title={Prompt}
]
\small
You are analyzing a complete Reddit discussion thread about AI-assisted software development using the supervisory workflow framework below. \\
\\
Identify every supervisory stage that is meaningfully discussed in the thread and extract all supporting evidence for each detected stage. \\
\\
Read the entire thread before making any decisions. \\
\\
GENERAL PRINCIPLE \\
\\
The stages represent supervisory functions in AI-assisted software-development work, not necessarily actions performed directly by the developer. \\
\\
A supervisory function may be performed by the developer or deliberately delegated to another agent, tool, or automated mechanism. Tag the stage when the developer intentionally uses that function to guide, oversee, coordinate, or evaluate AI-assisted work. \\
\\
Do not tag a stage merely because an AI agent autonomously performs a similar action while completing its task. \\
\\
CODING RULES \\
\\
- Tag a stage only when a concrete supervisory practice, experience, problem, failure, decision, or recommendation is explicitly described or clearly supported by the text. \\
- The evidence must concern supervising AI-assisted software-development work. General product-development practices, human-team management, or unrelated uses of AI do not count. \\
- General statements about AI capabilities, model quality, usefulness, reliability, or what someone could hypothetically do are not sufficient by themselves. \\
- Advice or recommendations may count when they describe a concrete supervisory practice, not merely a vague suggestion. \\
- A thread may contain multiple stages, and different participants may provide evidence for different stages. Do not combine statements from different participants into one continuous workflow. \\
- Do not force a stage to appear when the evidence is ambiguous. A thread may contain no supervisory stage at all. \\
- For each detected stage, include every independently stated example or discussion that provides meaningful evidence for that stage, even when similar practices are discussed by different participants or in different parts of the thread. Avoid duplicate or overlapping excerpts that capture the same statement from the same source. \\
- Each excerpt must be a contiguous, verbatim quote that includes enough context to understand why it supports the stage. Prefer complete sentences and avoid unnecessarily truncating the evidence. Do not paraphrase, combine, correct, or normalize the text. Copy source\_type, source\_id, and author exactly as supplied. \\
\\
STAGES \\
\\
- Plan: Structuring how an AI-assisted task should be carried out before an execution cycle by defining or refining requirements, constraints, task boundaries, success criteria, decomposition, or an intended approach. This may include using an agent to help develop or refine the plan. Simply naming a planning step, assigning a task, describing the current task, or giving general work advice does not count. \\
\\
- Monitor: Observing or checking an AI agent while a specific execution cycle is still ongoing to understand its progress, actions, tool use, intermediate state, errors, or goal drift. Merely knowing that an agent is running, describing general involvement, or tracking tokens, cost, latency, context usage, or other resource metrics does not count. \\
\\
- Wait: Allowing AI work to continue while the developer steps away, switches to another task, or works with another agent in parallel. The developer must leave the AI working without active intervention for some period. An agent waiting for developer input or approval, ordinary response latency, notifications, or remote-access functionality alone do not count. \\
\\
- Review: Evaluating AI-assisted work against the developer's goals, requirements, or quality criteria before accepting it or proceeding with subsequent work. Review may be performed directly by the developer or deliberately delegated to another agent, tests, static analysis, or other verification mechanisms. \\
\\
- Teach: Providing new information or direction after execution has begun in order to change the AI's subsequent work. This includes correcting misunderstandings, adding context, redirecting the AI, or asking it to revise, retry, or use another approach. Merely asking how feedback could be provided does not count. \\
\\
- Manual Fix: The developer directly modifies, debugs, corrects, completes, replaces, or works around AI-assisted work rather than delegating the correction back to the AI. \\
\\
- Update Assets: Creating or modifying a persistent resource intended to guide or improve future AI-assisted work, such as instructions, memory files, reusable prompts, project rules, documentation, tests, checklists, templates, or agent configurations. General discussion of improving a workflow or human-team rules does not count without a concrete persistent resource for AI-assisted work. \\
\\
KEY DISTINCTIONS \\
\\
- Plan structures how AI-assisted work will be carried out before an execution cycle; simply describing or assigning the task does not count. Teach changes the AI's direction after execution has begun. \\
- Monitor requires a concrete act of observing or checking an ongoing execution; Review evaluates work, results, or behavior that has already been produced. \\
- Wait means the developer leaves AI work running; an agent waiting for the developer is not Wait. \\
- Teach delegates a correction or change back to the AI; Manual Fix means the developer makes the correction directly. \\
- An agent performing planning, monitoring, testing, or reviewing on its own does not automatically constitute a supervisory stage. The function must be deliberately incorporated into the developer's supervisory process. \\
- Running tests or using another agent to evaluate current AI-assisted work may count as Review. Tests count as Update Assets when they are deliberately created or maintained as persistent guidance or guardrails for future AI work. \\
\\
OUTPUT \\
\\
Return only valid JSON: \\
\\
\{
  ``thread\_id'': ``\{\{THREAD\_ID\}\}'', \\
  ``has\_stage'': true, \\
  ``stage\_results'': [ \\
    \{ \\
      ``stage'': ``Plan | Monitor | Wait | Review | Teach | Manual Fix | Update Assets'', \\
      ``confidence'': ``high | mid | low'', \\
      ``excerpts'': [ \\
        \{ \\
          ``reason'': ``Brief reason this exact quote supports this stage'', \\
          ``source\_type'': ``post\_title | post\_body | comment'', \\
          ``source\_id'': ``Exact supplied source ID'', \\
          ``author'': ``Exact supplied author'', \\
          ``quote'': ``Exact contiguous verbatim quote'' \\
        \} \\
      ] \\
    \} \\
  ], \\
  ``reason\_if\_false'': ``'' \\
\} \\
\\
Include each detected stage at most once, with all supporting excerpts grouped under that stage. \\
\\
If has\_stage is true, stage\_results must contain at least one stage and reason\_if\_false must be an empty string. \\
\\
If no supervisory stage is supported, return: \\
\\
\{ \\
  ``thread\_id'': ``\{\{THREAD\_ID\}\}'', \\
  ``has\_stage'': false, \\
  ``stage\_results'': [], \\
  ``reason\_if\_false'': ``Brief reason no supervisory stage is supported'' \\
\} \\
\\
Do not return markdown or any text outside the JSON object. \\
\\
THREAD ID \\
\\
\{\{THREAD\_ID\}\} \\
\\
THREAD TO ANALYZE \\
\\
\{\{THREAD\}\} \\
\\
\end{tcolorbox}